%% file: 00_main.tex
\documentclass[11pt, letterpaper, authoryear]{article}
\usepackage{style/arxiv}

\RequirePackage[hyphens]{url}
\input{style/structure}

\title{\Large \textbf{Introducing ChatSQC: Enhancing Statistical Quality Control with Augmented AI}} %Allison's suggestion, feel free to edit/change

\setfinal

\date{\small \today}

\renewcommand{\shorttitle}{ChatSQC}

\begin{document}

\authorinfo 

\maketitle 

\begin{abstract}
\noindent We introduce ChatSQC, an innovative chatbot system that combines the power of OpenAI's Large Language Models (LLM) with a specific knowledge base in Statistical Quality Control (SQC). Our research focuses on enhancing LLMs using specific SQC references, shedding light on how data preprocessing parameters and LLM selection impact the quality of generated responses. By illustrating this process, we hope to motivate wider community engagement to refine LLM design and output appraisal techniques. We also highlight potential research opportunities within the SQC domain that can be facilitated by leveraging ChatSQC, thereby broadening the application spectrum of SQC. A primary goal of our work is to provide a template and proof-of-concept on how LLMs can be utilized by our community. To continuously improve ChatSQC, we ask the SQC community to provide feedback, highlight potential issues, request additional features, and/or contribute via pull requests through our public GitHub repository. Additionally, the team will continue to explore adding supplementary reference material that would further improve the contextual understanding of the chatbot. Overall, ChatSQC serves as a testament to the transformative potential of AI within SQC, and we hope it will spur further advancements in the integration of AI in this field.
\end{abstract}


%\thispagestyle{empty}

%\newpage

%\doublespacing

\clearpage

\section{Introduction}
\label{sec:intro}

\input{01_introduction}

\section{The Construction of ChatSQC} %\info{With better coding, more time, we should try to dive deeper into \href{https://arxiv.org/pdf/2304.05510v2.pdf}{chatClimate} for more functionality. Something for version 2.0.}}
\label{sec:methods}
\input{02_methods}

\section{Evaluation of ChatSQC}
\label{sec:experiments}
\input{03_experiments}

%\section{Results}
%\label{sec:results}
%\input{04_results}

\section{Discussion}
\label{sec:discussion}
\input{05_discussion}

\section{Concluding Remarks}
\label{sec:conc}
\input{06_conclusions}

\bibliographystyle{apalike}
\bibliography{refs, app_refs}

\appendix
\setcounter{table}{0}
\renewcommand{\thetable}{A\arabic{table}}

\section{Appendices}

\subsection{Code and GitHub Repository}
\label{app:code}
\input{07a_code}

\subsection{Materials for Interactive Verification}
\label{app:verifiability}
\input{07b_verifiability}

\subsection{Articles Used in the App}
\label{app:articles}
\input{07c_app_articles}
\bibliographystyleappx{apalike}
\bibliographyappx{refs, app_refs}

\subsection{Estimation of the ChatSQC App Usage Cost}
\label{app:cost}
\input{07d_cost}
\color{black}

\end{document}

%% file: style/structure.tex
\usepackage[utf8]{inputenc}

\usepackage[export]{adjustbox}
\usepackage{algorithm}
\usepackage[noend]{algpseudocode}
\usepackage{amsmath}
\usepackage{amssymb}
\usepackage{animate}
\usepackage{appendix}
\usepackage{array}
\usepackage{authblk}
\usepackage{bm}
\usepackage{booktabs}
\usepackage{caption}
\usepackage{color}
\usepackage{colortbl}
\usepackage{changepage}
\usepackage{chemformula}
\usepackage{comment}
\usepackage{csvsimple}
\usepackage{etoolbox}
\usepackage{enumitem}
\usepackage{float}
\usepackage{framed}
\usepackage{fontawesome5}
\usepackage{graphicx}
\definecolor{links}{HTML}{0074CC} % used to be 3366CC
\usepackage[colorlinks=true, allcolors=links]{hyperref}
\usepackage{lineno}
\usepackage{lscape}
\usepackage{listings}
\usepackage{makecell}
\usepackage{marginnote}
\usepackage{mathrsfs}
\usepackage{mathtools}
\usepackage{mdframed}
\usepackage{multicol}
\usepackage{multirow}
\usepackage[natbibapa]{apacite}
\usepackage{multibib}
\newcites{appx}{\small Open-Source Articles used to Augment ChatSQC-Research as of Version 1.3.0}
\usepackage{tabulary}
\usepackage{tikz}
\usepackage{setspace}
\usepackage{soul}
\usepackage{subcaption}
\usetikzlibrary{shapes, arrows, positioning, fit}
\usepackage{tabularray}
\usepackage{wrapfig}
\usepackage{xargs}
\usepackage[normalem]{ulem}
\graphicspath{ {figs/} }
\usepackage{epstopdf}
\PassOptionsToPackage{hyphens}{url}\usepackage{hyperref}

\definecolor{sandstorm}{rgb}{0.93, 0.84, 0.25}

 \bibpunct[, ]{(}{)}{,}{a}{}{,}%
\newcommand{\instructions}[1]{}

\newcolumntype{C}[1]{>{\centering\arraybackslash}p{#1}}

\providecommand{\keywords}[1]
{
  \noindent \small	
  \textbf{Keywords: } #1
}

\makeatletter
\newcommand*{\centerfloat}{%
  \parindent \z@
  \leftskip \z@ \@plus 1fil \@minus \textwidth
  \rightskip\leftskip
  \parfillskip \z@skip}
\makeatother

\AtBeginEnvironment{tabular}{\footnotesize} % Makes all tables at footnote size
\usepackage[colorinlistoftodos,prependcaption,textsize=tiny]{todonotes}
\newcommandx{\unsure}[2][1=]{\todo[linecolor=red,backgroundcolor=red!25,bordercolor=red,#1]{#2}}
\newcommandx{\change}[2][1=]{\todo[linecolor=blue,backgroundcolor=blue!25,bordercolor=blue,#1]{#2}}
\newcommandx{\info}[2][1=]{\todo[linecolor=orange,backgroundcolor=orange!25,bordercolor=orange,#1]{#2}}
\newcommandx{\improvement}[2][1=]{\todo[linecolor=sandstorm,backgroundcolor=	sandstorm!25,bordercolor=sandstorm,#1]{#2}}
\newcommandx{\thiswillnotshow}[2][1=]{\todo[disable,#1]{#2}}

\definecolor{miamired}{HTML}{C3142D}

\newif\ifblinded
\newif\iffinal

\newcommand{\setfinal}{
    \blindedfalse
    \finaltrue
}

\newcommand{\authorinfo}{
    \ifblinded
        \author{\large (Authors blinded for peer review)}
    \fi
    \iffinal
        \author[1]{Fadel M. Megahed}
        \author[2]{Ying-Ju Chen}
        \author[3]{Inez Zwetsloot}
        \author[4]{Sven Knoth}
        \author[5]{Douglas C. Montgomery}
        \author[1,*]{L. Allison Jones-Farmer}
        \affil[1]{Farmer School of Business, Miami University, Oxford, OH 45056, USA}
        \affil[2]{Department of Mathematics, University of Dayton, OH 45469, USA}
        \affil[3]{Amsterdam Business School, University of Amsterdam, Amsterdam, Netherlands}
        \affil[4]{Department of Mathematics \& Statistics, Helmut Schmidt University, Hamburg, Germany}
        \affil[5]{School of Computing and Augmented Intelligence, Arizona State University, Tempe, AZ 85281, USA}
        \affil[*]{Corresponding author can be reached at \href{mailto:farmerl2@miamioh.edu}{farmerl2@miamioh.edu}}
    \fi
}

%% file: 01_introduction.tex
Generative AI refers to a class of machine learning models that can generate text, images, and other content. Generative AI applications such as ChatGPT (with its current back-ends of GPT-3.5 and GPT-4 Turbo models), Gemini, Claude.ai, GitHub Copilot, Microsoft Copilot, Midjourney, and others have created worldwide excitement due to their ease of use, broad utility, and perceived capabilities. The conversational nature of some of these generative AI applications has reduced the barrier to entry, with no specialized software/hardware needed. OpenAI's ChatGPT amassed 1 million unique users within five days of its November 2022 launch and reached 100 million users about two months later. The use of AI is not limited to chat-based applications; for example, GitHub Copilot, which runs in coding interfaces (e.g., Visual Studio and \faRProject \ Studio), has been shown to increase developer productivity by more than 55\% \citep{peng2023impact} and satisfaction by 60-75\% \citep{kalliamvakou2022research}.

Focusing on \textit{text-to-text} generative AI models, two main trends have emerged in recent months. First, practitioners/researchers have increasingly examined the performance of these generally trained AI models in specific applications. For example, in our community, \citet{megahed2023generative} examined scenarios where ChatGPT (GPT-3.5) can inform Statistical Process Control (SPC) research, practice, and learning. In the general context of data science, \citet{shmueli2023can} examined possible (mis)uses of generative AI in the publication of \textit{data science} research. \citet{ellis2023new} examined how ChatGPT can enhance statistics and data science education. Second, large firms such as Google \citep{singhal2023large} and Bloomberg \citep{wu2023bloomberggpt} have developed specialized generative AI models to encode clinical and financial knowledge, respectively. These models have demonstrated superior performance compared to the then-current state-of-the-art general generative AI models.

The rapid adoption and profound impacts of generative AI are widespread throughout society, from science \citep{wang2023scientific} to labor productivity \citep{goldmansachs2023generative} and health care \citep{health2023embracing}. Despite its potential generative AI remains untapped within specialized fields such as SQC, where AI applications could transform research and practice. Pursuing such innovation has motivated us to develop \texttt{ChatSQC}, a ChatGPT-based tool provided with SQC context. Our approach is motivated by the work of \citet{vaghefi2023chatipcc}, who developed a \texttt{chatClimate} web app, available at \url{https://www.chatclimate.ai/}, which enhanced GPT-4's answers in the field of climate science. Hereafter, we use the term \texttt{SQC-augmented} to refer to that additional context. 

Recent literature reports significant advancements in augmenting existing large language models (LLMs) with domain-specific texts, a promising strategy for enhancing the precision and relevance of AI applications in niche areas. \cite{mialon2023augmented} conducted a survey on Augmented Language Models and emphasized the effectiveness of the missing token objective in enhancing reasoning capabilities. Similarly, \cite{peng2023check} proposed augmenting LLMs with external knowledge and automated feedback, significantly reducing hallucinations in ChatGPT. These studies underscore the importance of integrating additional information into LLMs to improve their performance. Furthermore, \cite{singhal2023large} focused on medical question answering with LLMs, showcasing progress towards achieving expert-level performance by leveraging base LLM improvements, domain-specific fine-tuning, and novel prompting strategies. In the context of orthopedic management, research, and patient queries, \cite{sosa2024capacity} explored the capacity of large language model chatbots to aid in these areas. Additionally, in addressing the grounding problem between cognitive and generative models, \cite{maher2023grounding} discussed ways to integrate these paradigms effectively, showing examples in computational creativity and education. The reader is referred to the surveys of \citet{minaee2024large}, \citet{gao2023retrieval} and \citet{tonmoy2024comprehensive} for detailed introductions on (a) the inception/motivation behind retrieval augmented generation (RAG), (b) the progression of RAG technologies, and (c) the utility of RAG methods on reducing hallucinations, respectively.

Our article demonstrates the first use case of how generative AI models can be augmented with SQC resources. We develop \texttt{ChatSQC}, a template and proof-of-concept on how our community can develop expertly grounded large language models. Our main contributions include the following.

%a unique perspective on how AI could be harnessed in the SQC field. Our objectives are:

\begin{enumerate}[nosep, label = (\arabic*)]
    \item \textbf{Augmenting a LLM with an SQC knowledge base:} We examine augmenting LLMs with further context by incorporating two types of high-quality SQC references (the \textit{NIST/SEMATECH e-Handbook of Engineering Statistics} and a corpus of all available open-source articles, with CC-BY and CC-BY-NC licenses, from leading journals in our field). We hypothesize that such context can enhance the LLMs' capabilities in providing domain-specific explanations and facts and that the parameters employed for reading data, separating the text, creating embeddings, and selecting the LLM will influence the output quality (presenting our community with a well-sized, yet manageable, playground to examine how our expertise can be used to enhance LLM design and output quality assessment).
    
    \item \textbf{Exploring opportunities for leveraging \texttt{ChatSQC} in SQC research:} Based on our findings, we intend to identify potential research avenues within SQC that could be pursued using \texttt{ChatSQC} and its future iterations.
    
    \item \textbf{Empowering practitioners with an SQC-augmented chatbot:} Our goal is to provide practitioners with a tool capable of delivering current/meaningful SQC-related responses that a general LLM may not generate, thus making cutting-edge SQC knowledge more accessible.
    
    \item \textbf{Crowdsourcing additional SQC knowledge bases for a better context:} To continuously improve the performance and relevancy of \texttt{ChatSQC}, we encourage the research and practice communities of SQC to contribute additional references via our unblinded GitHub repository: (fmegahed/chatsqc).
\end{enumerate}
By exploring the construction and application of \texttt{ChatSQC}, we hope to provide a pathway for practitioners to harness the potential of this tool and inspire researchers to continue pushing the boundaries of AI within the field of SQC.

%% file: 02_methods.tex
\begin{figure}[htb!]
    \centering
    \includegraphics[width = 0.98\textwidth, frame]{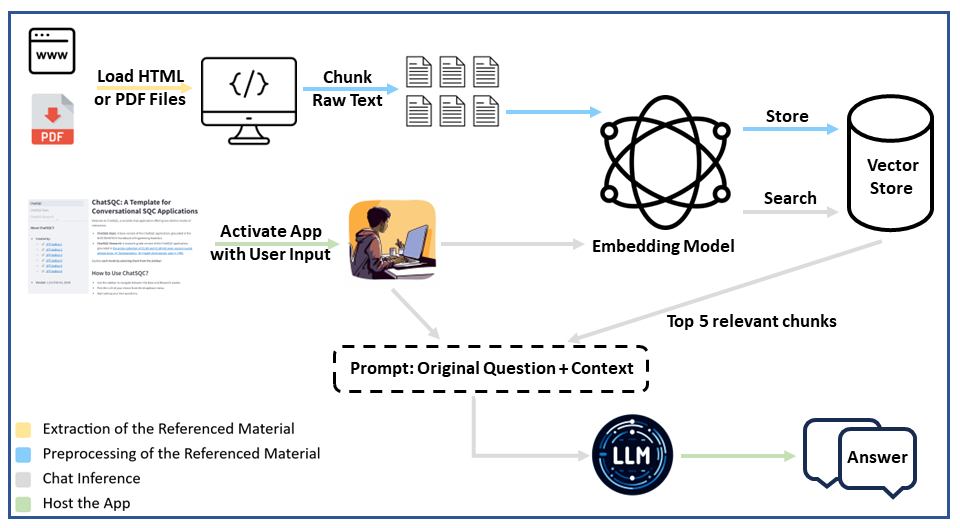}
    \caption{Flowchart of the preprocessing and chat interface creation process. The colors in the arrows correspond to the four phases highlighted in the paragraph above.}
    \label{fig:process_flowchart}
\end{figure}

The construction of \texttt{ChatSQC} involved four main phases: (a) a one-time extraction of the reference material\footnote{We will routinely scan the literature for CC-BY and CC-BY-NC papers that will be added to \texttt{ChatSQC-Research}. The one-time extraction refers to the status of the chatbot at the time of our second revision.}, (b) a one-time preprocessing of the extracted material, (c) a continuous (online) chat interface, and (d) the hosting/deployment of the app on a web server. An overview of our approach is depicted in Figure \ref{fig:process_flowchart}, with a detailed explanation below and our code presented in Section \ref{app:code}.

Our \texttt{ChatSQC} bot is designed to serve a wide range of users through its two specialized modes of interaction. The \texttt{ChatSQC-Basic} mode supports basic knowledge in SQC based on the concepts and techniques described in the \textit{NIST/SEMATECH e-Handbook of Statistical Methods}. The \texttt{ChatSQC-Research} mode provides access to the entire collection of CC-BY and CC-BY-NC open-access journal articles from \textit{Technometrics}, \textit{Quality Engineering}, and \textit{Quality and Reliability Engineering International (QREI)}, spanning from 2017 to January of 2024. No articles are included from \textit{Journal of Quality Technology (JQT)} and \textit{Quality Technology \& Quantitative Management (QTQM)} because none were published with the appropriate CC-BY or CC-BY-NC licenses; they were published with CC-BY-NC-ND which prevents derivative work such as the incorporation of these papers in our non-commercial chatbot.  The two modes of \texttt{ChatSQC} illustrate the application to the diverse needs of users by providing reliable, up-to-date information grounded in authoritative sources.

\subsection{Material Extraction and Integration Process}
\subsubsection{Extracting Reference Book Materials for ChatSQC-Basic}

 We chose the \textit{NIST/SEMATECH e-Handbook of Statistical Methods} as our reference book \citep{nist2022ehandbook} for \texttt{ChatSQC-Basic} since it was not marked as copyrighted.  The \citet{nist2021copyrights} site states, ``With the exception of material marked as copyrighted, information presented on NIST sites are considered public information and may be distributed or copied.'' The book comprises 694 web pages, capturing introductory material about industrial statistics, quality, reliability, and experimental design. 

We obtained the XML sitemap of the NIST E-handbook (\url{https://www.itl.nist.gov/div898/handbook/sitemap.xml}) and parsed it using the \texttt{Beautiful Soup} Python library \citep{richardson2023beautifulsoup}. Next, we reduced the 1,037 links from the NIST/SEMATECH E-handbook's XML sitemap to 694 URLs by keeping only the URLs associated with the different sections and subsections of the book. We sorted these 694 URLs to match the order by which the information is presented in the book. Finally, we used the \texttt{SeleniumURLLoader()} function from the \texttt{LangChain} \citep{langchain2023selenium} Python Library to read/load the text from these web pages and store it in a ``pickle file'', a Python object storage format that enables collected data to be saved in a serialized format that can be reloaded for future use. This allows for further processing without the need to extract information from the source each time.  

\subsubsection{Extracting Research Papers for ChatSQC-Research}

We used a collection of 52 open-access journal articles from (a) \textit{Technometrics}, (b) \textit{Quality Engineering}, and (c) \textit{QREI}, published between 2017 and 2024 with CC-BY and CC-BY-NC licenses.  These licenses allow us to create derivative work (as would be synthesized by an LLM) for non-commercial purposes. The list of articles used in the app can be found in the Appendix \ref{app:articles}.\footnote{The list includes papers used as of the end of January 2024. The papers used to ground \texttt{ChatSQC} will be updated and cited on the app's landing page.} The \cite{creativecommons2023cclicenses} indicates that both licenses grant users to distribute, remix, adapt, and build upon the original material in any medium or format, with the requirement that credit must be given to the creator. The key distinction between the two lies in their approach to commercial use: the CC-BY license permits commercial use of the material, while the CC-BY-NC license restricts use to noncommercial purposes only. 

To identify papers published by \textit{Taylor \& Francis} with the correct open access licenses, we combined the CrossRef API (\url{https://www.crossref.org/documentation/retrieve-metadata/rest-api/}) and the published list of open-access papers for each journal (e.g., see \url{https://www.tandfonline.com/action/showOpenAccess?journalCode=utch20} for \textit{Technometrics}).  For papers published in \textit{QREI}, we were provided a CSV file (\url{https://github.com/fmegahed/chatsqc/blob/main/data/qrei_open_access_list_wiley.csv}) from \textit{Wiley} to identify all open-access papers with the correct license. For all journals considered, we manually downloaded the 52 papers (PDF files). We stored them in a directory to be preprocessed and read by our Python code, which we created a citation dictionary for all our papers saved as a pickle file and a CSV file with the citation information for all papers.

%In prototyping \texttt{ChatSQC}, we chose the \textit{NIST/SEMATECH e-Handbook of Statistical Methods} as our reference book \citep{nist2022ehandbook}. We chose that book since the book was not marked as copyrighted; per the \citet{nist2021copyrights} site: ``With the exception of material marked as copyrighted, information presented on NIST sites are considered public information and may be distributed or copied.'' Furthermore, we estimate that the book comprises 694 web pages that capture introductory material about industrial statistics, quality, reliability, and experimental design. Hence, we believe using this public-domain resource would suit our prototype version of \texttt{ChatSQC}. 

%The extraction and merging of the different web pages consisted of several consecutive steps. First, we obtained the XML sitemap of the NIST E-handbook (\url{https://www.itl.nist.gov/div898/handbook/sitemap.xml}), and parsed it using the \texttt{Beautiful Soup} Python library \citep{richardson2023beautifulsoup}. In the second step, we reduced the 1,037 links from the NIST/SEMATECH E-handbook's XML sitemap to 694 URLs by keeping only the URLs associated with the different sections and subsections of the book. Third, we sorted these 694 URLs to match the order by which the information is presented in the book. Fourth, we used the \texttt{SeleniumURLLoader()} function from the \texttt{LangChain} \citep{langchain2023selenium} Python Library to read/load the text from these web pages and store it in a pickle file for further text preprocessing and analysis. 

\subsection{Offline Preprocessing}

For \texttt{ChatSQC-Basic}, we extracted the raw text from the pickle file using the \texttt{get\_pickle\_text()} function, obtaining a single string containing the entire reference book content. Similarly, for \texttt{ChatSQC-Research}, the \texttt{DirectoryLoader} and \texttt{PyMuPDFLoader} Python methods from the \texttt{langchain} package were used to read the PDF files. We divided the large text into smaller, more manageable chunks using the \texttt{LangChain} library's \texttt{RecursiveCharacterTextSplitter()} function. For this version of \texttt{ChatSQC}, we split the text into chunks of a maximum of 1,000 characters with a 10\% overlap to ensure contextual coherence. This choice is a balanced approach that carefully considers the trade-offs between contextual coherence, computational efficiency, and the practical limits of embedding models. Large chunks could capture more context but would increase memory usage and computational time and potentially introduce more noise. Conversely, smaller chunks could miss important contextual information, leading to poorer-quality embeddings. Increasing the overlap can enhance context preservation between chunks but result in redundant processing, which can be computationally inefficient. Reducing the overlap might make the process more efficient but at the risk of losing important transitional context, affecting the coherence of the information processed across chunks. The 10\% overlap is a practical choice, providing context continuity without excessively increasing the computational load. The choice of the maximum number of characters and the overlap size are parameters that can be tuned in future versions of \texttt{ChatSQC}.

Next, we have originally used OpenAI's \texttt{text-embedding-ada-002} embedding model to vectorize each data chunk. Text embeddings allow us to assess the similarity between text strings and find widespread use in various natural language processing (NLP) tasks like searching and ranking. Text embeddings represent text as sets of floating-point number vectors, and the distance between these vectors reflects their similarity or connection. Smaller distances indicate higher relatedness, whereas larger distances imply lower relatedness.
%Text embeddings allow us to measure the relatedness of different text strings and are widely used in various natural language processing (NLP) tasks, including search and ranking based on relevance to a query string. These embeddings represent text as vectors of floating-point numbers, where the distance between two vectors indicates their similarity or relatedness. Smaller distances imply higher relatedness, while larger distances suggest lower relatedness. \info{Previous two sentences are from \url{https://platform.openai.com/docs/guides/embeddings/what-are-embeddings} and we will need to paraphrase them} 
One of the key features of \texttt{text-embedding-ada-002} is its tokenization process, which breaks down text into manageable units, or tokens, for analysis. Unlike simple tokenization methods that might only split text on spaces or punctuation, the model employs a more sophisticated approach that can handle a variety of languages and linguistic nuances. A token can represent a whole word, part of a word, or even multiple words, depending on the tokenizer's granularity. This allows the model to efficiently process and understand text ranging from short sentences to longer documents. We initially chose \texttt{OpenAI's text-embedding-ada-002} because it was trained on a diverse dataset, encompassing a wide range of languages and domains. This extensive training enables the model to generate more refined and accurate embeddings to capture textual similarity and relatedness. Compared to other embedding models at the time of the initial analysis, \texttt{text-embedding-ada-002} was a robust, cost-effective, and was OpenAI's most powerful embedding model. We can access this embedding model through OpenAI's API at a price of \$0.0001 per 1,000 tokens (a token can be a word or a part of a word depending on the choice of the tokenizer) \citep{openai2023embeddings}. Notably, in English, approximately 1,000 tokens are roughly equivalent to 750 words, providing a useful guideline for estimating processing costs. We estimate the costs of embedding all our pre-referenced text to be less than \$1 (See Appendix \ref{cost:emb}). For a detailed introduction to text embeddings, we refer the reader to \citet{openai2023embeddings} and \citet{langchain2023text}. 

Finally, we created a vector-based database that can be used to store embeddings of the text chunks and their metadata (i.e., the URLs of different text chunks for \texttt{ChatSQC-Basic} or the URLs of journal papers and their licenses for \texttt{ChatSQC-Research}), allowing us to efficiently retrieve the most relevant documents based on their similarity to a specific query's embedding. We used the FAISS (Facebook AI Similarity Search) vector store in our prototype to implement this due to its robustness and efficiency in handling large-scale similarity searches \citep{douze2024faiss}. By using FAISS, we can store this offline preprocessing database locally, eliminating the need for additional internal or external servers \citep{langchain2023vs}. 
% This practical feature streamlines the implementation process and ensures seamless data retrieval \citep{langchain2023vs}.

The reader should note that our app has been updated (i.e., we repeated the preprocessing of all referenced materials) to the current-state-of-the-art OpenAI embeddings model \texttt{text-embedding-3-large} since it performs best on evaluation benchmark datasets. For \texttt{ChatSQC}, the cost of embedding the referenced materials remains under \$1 with our newly chosen embeddings model; \citet{openai2023embeddings} estimates that \$1 can embed 9,615 text pages based on their assumption of 800 tokens/page).\footnote{Due to the rapidly evolving nature of LLM developments, we will continue to update our app/chatbot with new LLMs, embeddings, and SQC-referenced materials to continue to improve our chatbot's performance and its relevancy to our community. Given that we use GitHub to capture code changes, interested researchers can use our revision history to go back to a specific ``as-of'' version of our bot.} 

\subsection{The Online Chatting Interface}
\label{sec:interface}
The online chatting interface of our \texttt{ChatSQC} prototype facilitates seamless interactions between users and the application. It comprises several key components that work together to enable a dynamic and contextually-aware conversation experience. First, we constructed a conversation chain to capture the sequence of exchanges between the user and our application, preserving the context of ongoing conversations (from a chat history and a grounding in our reference materials perspective). The conversation chain integrates three main components:
\begin{enumerate}[label=(\alph*), nosep]
\item \textbf{ChatOpenAI:} We have initially used OpenAI's \texttt{gpt-3.5-turbo-16k} to generate responses to user inputs. The GPT-3.5 model has been known for its efficient text generation capabilities with a context window of 16,385 tokens, offering a balance between performance and computational efficiency \citep{gpt35_16k_2023}. It was designed to handle various natural language processing tasks, including conversation, text completion, summarization, and more, with faster response times and lower costs than earlier versions. The GPT-3.5 model was used to understand user questions and provide relevant answers based on the information available in the reference book. However, as we revised this manuscript, a more powerful GPT-4 Turbo model (i.e., \texttt{gpt-4-turbo-preview}) was available, and it is now used for both the \texttt{ChatSQC-Basic} and \texttt{ChatSQC-Research} modes of our app. The prompt responses for the \texttt{ChatSQC-Research} mode, shared in this manuscript, utilize the latest model. Note that the GPT-4 Turbo model has a context window of 128,000 tokens and has training data up to December 2023.
\item \textbf{Vector Store as a Retriever:} To enhance the relevance and accuracy of responses, \texttt{ChatSQC} employs a vector retriever component, which retrieves the five most relevant chunks of text from a preprocessed dataset, based on the user's input. This dataset contains metadata for individual web pages or text chunks, serving as a critical resource for providing insightful evidence to users. The retrieval process operates through vector similarity comparison: When a user inputs a query, this component translates the query into a vector embedding and then searches the dataset for the top five text chunks whose embeddings most closely match the query's embedding. This process relies on advanced vector space modeling to understand the semantic nuances of the user's input beyond mere keyword matching.

\item \textbf{RAG:} After retrieving the top 5 relevant text chunks, the \texttt{RAG} component dynamically integrates this retrieved information into the response generation process of the LLM. By augmenting the generative model's input with contextually rich, retrieved text, RAG significantly enhances the quality and relevance of the generated responses, ensuring they are informed by the most pertinent information available.
%These text chunks and their corresponding metadata are archived in the vector store. This information repository holds individual webpage or text chunk metadata and serves as a critical resource for providing insightful evidence to our users. This retrieval process is determined by the similarity between the user's input embedding and our dataset's embeddings of text chunks. 
\item \textbf{Memory Buffer:} The conversation chain is augmented with a \texttt{ConversationBufferMemory} object, which maintains the chat history. This buffer stores the sequence of user inputs and model responses, essential for generating coherent/ contextually appropriate answers.
\end{enumerate}
We integrated a custom system message prompt (see Figure \ref{fig:system_prompt}) corresponding to the \texttt{ChatSQC} mode into the conversation to provide responses in adherence to specific guidelines and avoid relying on ChatGPT's in-house knowledge. The custom prompt enhances transparency and ensures that users receive answers solely based on the information provided by the user and the reference book.

\begin{figure}[htb!]
\begin{subfigure}{\textwidth}
\centering
\begin{minipage}{\linewidth}
\begin{lstlisting}[language=Python]
system_message_prompt = SystemMessagePromptTemplate.from_template(
    "You are a Q&A bot, an intelligent system that answers user questions ONLY based on the information provided by the user. When you use the information provided by the user, please include `\\n (Source: NIST/SEMATECH e-Handbook of Statistical Methods)' at the end of your response with a line break. If the information cannot be found in the user information, please say, `As a SQC chatbot grounded only in NIST/SEMATECH's Engineering Statistics Handbook, I do not know the answer to this question as it is not in my referenced/grounding material. I am sorry for not being able to help.' No answers should be made based on your in-house knowledge. For example, you may know what a large language model is, but that information does not come from the knowledge base that we provided to you. So defining a large language model based on your knowledge is unacceptable. Obviously, other algorithms, descriptions, and formulas that are not in the knowledge base we provided are also unacceptable. The context is:\\n{context}."
)
\end{lstlisting}
\end{minipage}
\vspace{-0.5\baselineskip}
\caption{\texttt{ChatSQC-Basic's} system prompt.}
\end{subfigure}
\vspace{1em}
\begin{subfigure}{\textwidth}
\centering
\begin{minipage}{\linewidth}
\begin{lstlisting}[language=Python]
system_message_prompt = SystemMessagePromptTemplate.from_template(
      "You are a Q&A bot, an intelligent system that answers user questions ONLY based on the information provided by the user. If the information cannot be found in the user information, please say 'As a SQC chatbot grounded only in open-access SQC research papers, I do not know the answer to this question as it is not in my referenced/grounding material. I am sorry for not being able to help.' No answers should be made based on your in-house knowledge. For example, you may know what a large language model is, but that information does not come from the knowledge base that we provided to you. So defining a large language model based on your knowledge is unacceptable. Obviously, other algorithms, descriptions, and formulas that are not in the knowledge base we provided are also unacceptable. The context is:\n{context}."
)
\end{lstlisting}
\end{minipage}
\vspace{-.5\baselineskip}
\caption{\texttt{ChatSQC-Research's} system prompt.}
\end{subfigure}

\vspace{-.65\baselineskip}

\caption{Our custom system prompts for the two modes of \texttt{ChatSQC}.}
\label{fig:system_prompt}
\end{figure}

Next, we developed a function, \texttt{handle\_userinput()},  to process the user's questions and prompt the appropriate responses. This function receives the user's input and uses the conversation chain to generate a LLM response. The response is then appended to the chat history stored in the session state. The session state, a feature of the Streamlit framework, allows data to persist for a given user-session. We added three features to this function: (a) a document similarity search, where we leverage our vector retriever to show up to five of the most similar chunks of text in our preprocessed dataset and present them to the user for reference; (b) the $L2$ distance between the user's prompt and the shown chunks of text; and (c) the incorporation of the webpage title, from which the text chunk originates, hyperlinked to the actual URL for quick access. %In our generation of the query, the five most relevant chunks of text are used to formulate a response; however, we display the closest two for brevity. 

\begin{figure}
    \centering
    \includegraphics[height=0.925\textheight, frame]{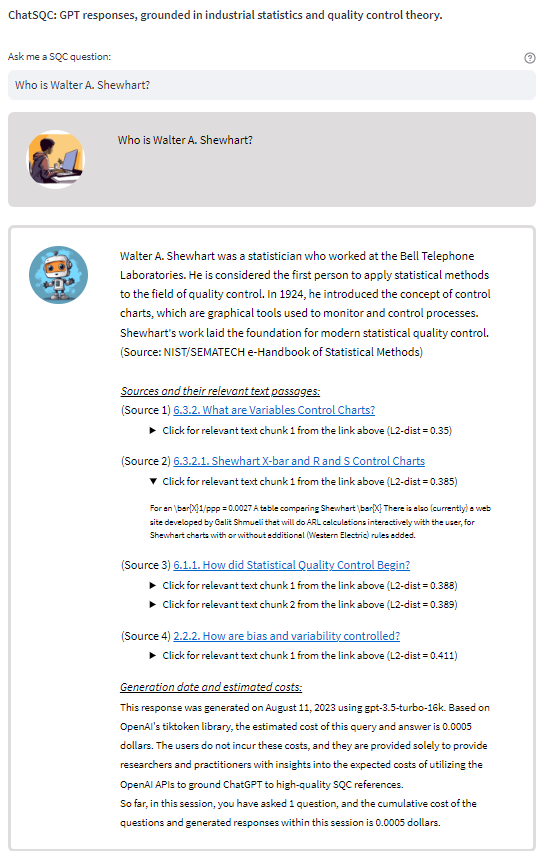}
    \caption{Screenshot of \texttt{ChatSQC-Basic} where the used prompted the question ``Who is Walter A. Shewhart?'' and its response. This screenshot is only provided to highlight the functionalities of our app and how it provides some features that are not available by default in (most) existing Chat models.}
    \label{fig:chatsqc-screenshot}
\end{figure}

To orchestrate the interaction flow of the app, we designed a \texttt{main()} function, which prompts the user for a question and forwards it to the \texttt{handle\_userinput()} function for processing and generating appropriate responses. Figure \ref{fig:chatsqc-screenshot} shows a screenshot of \texttt{ChatSQC}. The user asked the question, ``Who is Walter A. Shewhart?'' and received an appropriate answer grounded in the NIST/SEMATECH E-handbook. Following \texttt{ChatSQC}'s answer, links to five similar chunks of text in our preprocessed dataset are provided with the $L2$ distance between the user's prompt and a similar chunk. Displaying L2 distances enhances transparency and user trust by providing a clear, quantifiable measure of relevance. It allows users to quickly assess the similarity of search results or recommendations, offering insights into the model's perception of similarity and aiding in refining queries for more precise outcomes. Finally, information about the response generation date and estimated query costs are offered. Knowing when the responses are generated and understanding the costs of queries can be crucial in various contexts. Organizations can ensure that data are current and aligned with regulatory standards; they can manage and control budgets, optimize queries to reduce expenses, plan and forecast financial commitments, and avoid overuse of the app, etc.

%The user has the option to click and expand the white box below the question box to inspect which two (out of five) chunks of text in our preprocessed dataset where used for generating the response (Figure \ref{fig:textchunks}). The first chunk of text (Figure \ref{fig:textchunks}a) comes from Chapter 6.3.2. entitled ``What are Variables Control Charts?" \footnote{https://www.itl.nist.gov/div898/handbook/pmc/section3/pmc32.htm} and the second chunk of text (Figure \ref{fig:textchunks}b) comes from Chapter 6.1.1. entitled ``How did Statistical Quality Control Begin?" \footnote{https://www.itl.nist.gov/div898/handbook/pmc/section1/pmc11.htm}

\subsection{The Hosting/Deployment of ChatSQC}

The app was containerized and deployed by the Ohio Supercomputer Center (OSC). The OSC is a prominent facility for high-performance computing and an essential resource for academic, industrial, and governmental users in the State of Ohio. The OSC server delivers a stable implementation, accommodating a substantial number of users for \texttt{ChatSQC}. We host the app on \url{https://chatsqc.osc.edu/}. The project is stored in a GitHub repository, and the OSC administrators automatically update the published app when commits are made.

%% file: 03_experiments.tex
To evaluate the information generated by our \texttt{ChatSQC} app, we have performed a comparative study with two benchmark methods, 
\texttt{GPT-3.5} and \texttt{GPT-4}. \texttt{GPT-4} was run within \texttt{ChatGPT plus}, a subscription version of \texttt{ChatGPT}. Hereafter, we will refer to these three LLMs as \texttt{ChatSQC}, \texttt{GPT-3.5}, and \texttt{GPT-4}. We compare both \texttt{ChatSQC-Basic} and \texttt{ChatSQC-Research} to the benchmarks. Our goal in evaluating \texttt{ChatSQC} is to illustrate that it provides grounded responses, guardrails against false information and, in some cases, it can do things that the state-of-the-art \texttt{GPT-4} model does not do well. Our app is currently in its \textbf{Alpha stage}, presented as an initial tool to ground generative AI models in high-quality SQC references. \footnote{For the comparison with \texttt{ChatSQC-Basic} we used a customized \texttt{GPT-3.5}. The customized \texttt{GPT-3.5} utilizes the API model titled ``chatgpt-3.5-turbo-16k", which OpenAI plans to depreciate on June 13, 2024 (users can opt to use \texttt{gpt-3.5-turbo} which points to the most up to date version of the GPT-3.5 Turbo models). For the comparison with \texttt{ChatSQC-Research} we used the online available web versions of \texttt{ChatGPT-3.5 and ChatGPT-4 Turbo} in February 2024.}

\subsection{Study design}

Following a similar approach to \citet{megahed2023generative}, we have used a structured approach that compares responses from each platform using expert raters. As a first step in our evaluation approach, we have designed prompts (shown in Table \ref{tab:prompts}). Prompts 1-8 were used to compare \texttt{ChatSQC-Basic} with the benchmarks and are designed to capture the following subfields of SQC: monitoring, measurement, reliability, and experimentation. Note that (a) the prompts have been labeled by their type (explanation, application, evaluation); (b) the numbering captures the order by which these prompts were entered into the chatbots and not the different types. Prompts 9 and 10 were only evaluated in \texttt{ChatSQC} (and not in the benchmarks) as these prompts have been included to showcase how \texttt{ChatSQC} deals with prompts for which the answers are not included in the reference material. Prompt 11 was designed to compare \texttt{ChatSQC-Research} with the benchmarks. We selected 11 papers from the reference material to insert into prompt 11. In Table \ref{app:articles}, we have indicated these papers with an asterisk.

\begin{table}
    \centering
    \caption{Prompts used in the study and their characteristics}
    \label{tab:prompts}
    \begin{tabular}{cm{2.5in}C{0.7in}C{0.7in}C{0.7in}} \toprule
    &\textbf{Prompt}&\texttt{\textbf{ChatSQC}} \textbf{mode}&\textbf{Type} &\textbf{Contained in ref. material}\\
    \toprule
    \textbf{Prompt 1} & Can you explain the difference between phase 1 and phase 2 control charting applications in statistical process monitoring?&Basic&Explanation  &Yes \\
    \midrule
    \textbf{Prompt 2} & When should I use a univariate and a multivariate monitoring approach? &Basic& Application &Yes\\
    \midrule
    \textbf{Prompt 3} & How do I differentiate between fixed and random factors for measurement uncertainty quantification?&Basic&Evaluation&Partially\\ \midrule
    \textbf{Prompt 4} & Explain the difference between repeatability and reproducibility in a Gage R\&R study?&Basic&Explanation&Yes\\ \midrule
    \textbf{Prompt 5} & What steps should I take when performing a reliability assessment?&Basic&Evaluation&Partially\\ \midrule
    \textbf{Prompt 6} & How can you estimate distributional parameters from censored data?&Basic&Application &Yes\\ \hline
    \textbf{Prompt 7} & Is it a problem that my factor is aliased with a 3-way interaction in my design?&Basic&Evaluation&Partially\\ \midrule
    \textbf{Prompt 8} & What does D-optimal mean in the context of experimental design? &Basic&Explanation  &Yes\\     \midrule
    \textbf{Prompt 9} & %Who is Philip B. Crosby?
    What is a synthetic control chart? & Basic and Research&Explanation  &No\\ \midrule
    \textbf{Prompt 10} & Can you define temperature? &Basic and Research&Explanation  &No \\ \midrule
    \textbf{Prompt 11} & Summarize the key findings in [insert ref to paper] &Research &Explanation  &Yes \\
    \bottomrule      \end{tabular}

\end{table}

In the second step, responses from \texttt{ChatSQC} (Basic or Research), \texttt{GPT-3.5}, and \texttt{GPT-4} were generated independently by inputting each prompt into a fresh session of the app. Once the response was recorded, the session was refreshed by either reloading the web app (\texttt{ChatSQC}, \texttt{mGPT3.5}) or by starting a new chat (\texttt{ChatGPT-3.5} and \texttt{GPT-4}) before entering the next prompt. While the prompts were entered in the same sequential order in each app, this should not impact the generated responses since each session was refreshed/reloaded before entering the next prompt. One of the paper's authors, who did not participate in evaluating the responses, prompted the apps and recorded the responses. 

In the third step, the responses of the LLMs were rated on accuracy, using a five-point Likert scale with responses ranging from 1=``Entirely Inaccurate'' to 5=``Entirely Accurate'', with a possible NA rating if the LLM indicated that the information was not available. The rubric used for the evaluation is depicted in Table \ref{tab:evaluation_rubic}.

\begin{table}
    \centering
    \caption{The scale used for evaluating the LLM responses. Within [square brackets], the description used from prompts 1-8, and within (round brackets), the description used for prompt 11.}
    \label{tab:evaluation_rubic}
    \begin{tabular}{ccm{4.7in}} \toprule
     \textbf{Rating} &  \textbf{Short Description}    & \textbf{Detailed Description} \\ \midrule
     NA     &  Not Applicable & The response indicated that it does not have [the information in the source files] (access to the paper/knows the paper).\\ \midrule
      1      &  Entirely Inaccurate  & The answer is entirely inaccurate, demonstrating a lack of understanding or severe misconceptions about the [concepts/questions] (paper) involved.\\ \midrule 
     2      &  Partially Accurate   & The answer shows some understanding of the [topic] (paper) but contains major inaccuracies or omissions that significantly compromise the correctness and/or completeness of the answer. \\ \midrule 
     3      &  Generally Accurate   & The answer is generally correct but lacks full depth and/or precision. The answer may contain minor inaccuracies that do not significantly detract from the overall answer. \\ \midrule 
     4      &  Mostly Accurate      & The answer is mostly correct, showing a high level of understanding, but may lack the finer details or show tiny inaccuracies. \\ \midrule 
     5      &  Entirely Accurate    & The answer is entirely accurate, displaying a thorough understanding of the [concept] (paper), precision in explanation, and all necessary details are accurate. \\ \bottomrule
    \end{tabular}
\end{table}

The fourth step is different for the comparison of \texttt{ChatSQC-Basic} with the benchmarks and \texttt{ChatSQC-Research} with the benchmarks. To evaluate \texttt{ChatSQC-Basic}, we used the responses to prompts 1-8. The responses were blinded (by the author who prompted the three chatbots) so that the raters would not know which app generated the responses. The order of presentation of the responses was also randomized to prevent the raters from guessing the source. Four authors, experts in SQC, rated each response on accuracy. To evaluate \texttt{ChatSQC-Research}, we use the responses to prompt 11, which we repeat for 11 papers. We have asked the corresponding author of each paper to evaluate the responses on accuracy. We received replies from 7 out of 11 of these corresponding authors. We have excluded any papers written by the authors of this manuscript from this comparison. 

It is important to realize that there is inherent variability in the responses provided by LLMs. LLMs typically have a tuning parameter called temperature, which can be used to control the amount of randomness in the response. Understanding how this randomness affects the usefulness and correctness of the response is outside the scope of this research and the topic of future study. In both ChatSQC modes, we set the temperature parameter to 0.25 on a 0 (most consistent) to 2 (most creative) scale \citep{openai2024temp}. Note we cannot directly control the temperature parameter on the ChatGPT web interface. However, we provide all generated responses in the supplementary online materials (see Section \ref{app:verifiability} for CSV files of the expert accuracy ratings and links to screenshots of the responses and output).

\subsection{\texttt{ChatSQC-Basic} Responses}
%Talk briefly about the results. The focus should be on the potential (and beauty of having control) and the emphasis that our app is in its Alpha stage. 
In this subsection, we focus on evaluating \texttt{ChatSQC-Basic} by presenting the ratings for the responses to prompts 1-8 generated by the three apps.

Table \ref{tab:descriptives} presents the average accuracy ratings and the standard deviation for each of the three apps, averaged across the first eight prompts and four raters ($n=32)$ as well as the average word count and the standard deviation for each of the three apps averaged across the first eight prompts ($n=8$). The results suggest that \texttt{ChatSQC-Basic} and \texttt{GPT-3.5} have very similar accuracy values. As expected, \texttt{GPT-4's} accuracy is slightly higher, in line with the well-documented belief that \texttt{GPT-4} provides more accurate responses than \texttt{GPT-3.5} \citep{openai2023gpt4}. We have also included the average word count and see that our \texttt{ChatSQC-Basic} provides much shorter answers, on average, than \texttt{GPT-3.5} and \texttt{GPT-4}. We see that GPT-4 provides about twice as long answers as our \texttt{ChatSQC-Basic}.

It should be noted that our app can also provide responses based on the \texttt{GPT-4} API. We did not include them in the evaluation since this feature was added after the raters were provided with the three apps' prompt responses. In our hosted version of the \texttt{ChatSQC} app, we allow users to generate responses using either the ``chatgpt-3.5-turbo-0125'' or ``chatgpt-4-turbo'' APIs.\footnote{At the time of our first submission, we used the ``chatgpt-3.5-turbo-16k'' and ``chatgpt-4'' models, then-state-of-the-art models.}

\begin{table}[htb!]
    \centering
    \caption{Descriptive statistics of the ratings for prompts 1-8.}
    \label{tab:descriptives}
    \begin{tabular}{c ccc } 
    \toprule
     & \textbf{\texttt{ChatSQC-Basic}} & \textbf{GPT-3.5}& \textbf{GPT-4}\\ 
    \midrule
    Average accuracy (std dev) & 4.25 (1.05) & 4.28 (0.81)& 4.44 (0.98)\\
    Average word count (std dev) & 198.1 (61.3) &  259.4 ( 98.9)  & 405.9 (78.4) \\
    $n$ accuracy ratings       & 32   & 32   & 32\\
    $n$ prompts      & 8   & 8  & 8\\
    \bottomrule      
    \end{tabular}
\end{table}

In Table \ref{tab:rating}, we present the average accuracy ratings for each prompt averaged across the four raters. Note that our presentation follows the prompt type and not the order by which the prompts were entered into the system. In the paragraphs below, we discuss the obtained results.

\begin{table}[htb!]
    \centering
    \caption{The average accuracy obtained from the four expert raters split by prompt.}
    \label{tab:rating}
    \begin{tabular}{c ccc cc} \toprule
    \textbf{Prompt} & \textbf{ChatSQC} & \textbf{GPT-3.5} & \textbf{GPT-4 }& Prompt type&Contained in handbook\\ 
    \midrule
    1  &  5    &  5    &  5\\
    4  &  5    &  5    &  5    &  Explanation &Yes\\
    8  &  5    &  4.25 &  4.5\\
    \midrule
    2  &  5	   &  4	   &  3.75 &  Application & Yes\\
    6  &  3.25 &  4.25 &  5 \\
    \midrule
    3  &  3.5  &  3.5  &  3.75\\
    5  &  4.25 &  4.5  &  4.25 &  Evaluation &Partially\\
    7  &  3	   &  3.75 &  4.25\\

    \bottomrule      
    \end{tabular}
\end{table}

First, let us consider the results for the explanation-type prompts (prompts 1, 4, and 8) for which the handbook includes the answer. Here, we see that our \texttt{ChatSQC-Basic} provided perfect responses and outperforms the benchmark methods slightly for prompt 8 (``What does D-optimal mean in the context of experimental design?''). This is unsurprising as prompt 8 is a specialized topic for which general LLMs like \texttt{GPT} are more likely to have generic answers. In contrast, our app, grounded in SQC references, will respond more precisely (as long as the topic is covered sufficiently in our referenced book). 

Second, let us consider the two application-type prompts (2 and 6) for which the handbook also contains the answer. In prompt 2, we asked ``When should I use a univariate and a multivariate monitoring approach?" This question relates to the choice of control charting approaches in statistical process monitoring, and receives higher scores from the raters for the \texttt{ChatSQC-Basic} response than the responses generated by \texttt{GPT-3.5} and \texttt{GPT-4}. On the other hand, prompt 6 relates to a broader knowledge of reliability, asking, ``How can you estimate distributional parameters from censored data?". The handbook provides two methods to do this, and \texttt{ChatSQC-Basic} accurately and clearly states these two in its response; however, this response receives a low accuracy score by the raters as more methods exist. The benchmark methods mention a total of 4 (\texttt{GPT-3.5}) and 6 (\texttt{GPT-4}) estimation methods. In the case of prompt 6, the accuracy reflects the limitation of the handbook's content since \texttt{ChatSQC} synthesized the information based on the handbook's description. Adding more detailed sources to our \texttt{ChatSQC} could improve its performance for such a question. 

Third, let us examine \texttt{ChatSQC-Basic's} answer to prompts 3, 5, and 7, which are partially contained in the handbook. As expected, \texttt{ChatSQC-Basic's} response accuracy to these prompts is lower than that of \texttt{GPT-3.5} and \texttt{GPT-4}, which were fully described in our reference material. If we examine all three apps, we note that: (a) none of the apps received an average score of 5 (entirely accurate) for these three prompts, and (b) the average scores varied between 3 (generally accurate) to 4.5 (mostly accurate) for the three apps. We expect that our \texttt{ChatSQC-Basic} would improve the accuracy of its responses if more SQC sources would be included in the knowledge base. Of course, this improvement may not be possible for the benchmark methods since their training has been completed, and OpenAI controls the training of future versions of GPT. 

\subsection{Responses to Ungrounded Prompts}

\begin{figure}[htb!]
    \centering
     \begin{subfigure}[T]{0.495\textwidth}
         \centering
         \includegraphics[width=\textwidth, trim={0 0 0 1.3in}, clip]{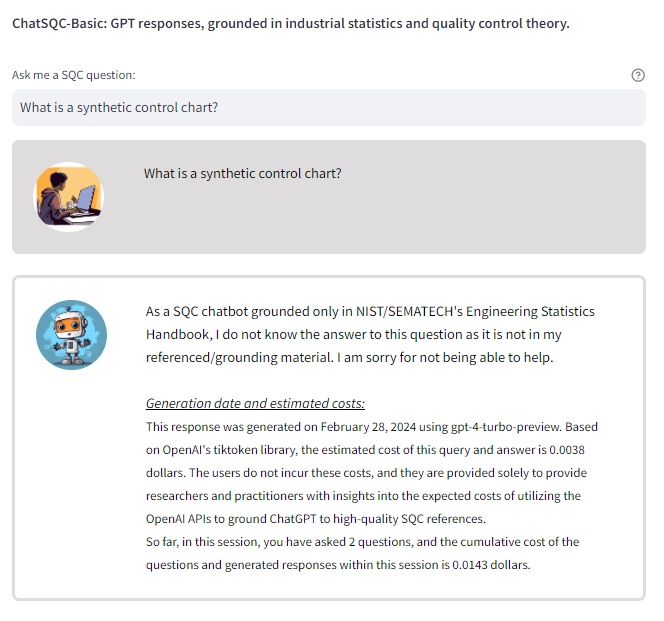}
         \subcaption[a]{\texttt{ChatSQC-Basic}}
     \end{subfigure}
          \begin{subfigure}[T]{0.495\textwidth}
         \centering
         \includegraphics[width=\textwidth, trim={0 0 0 .8in}, clip]{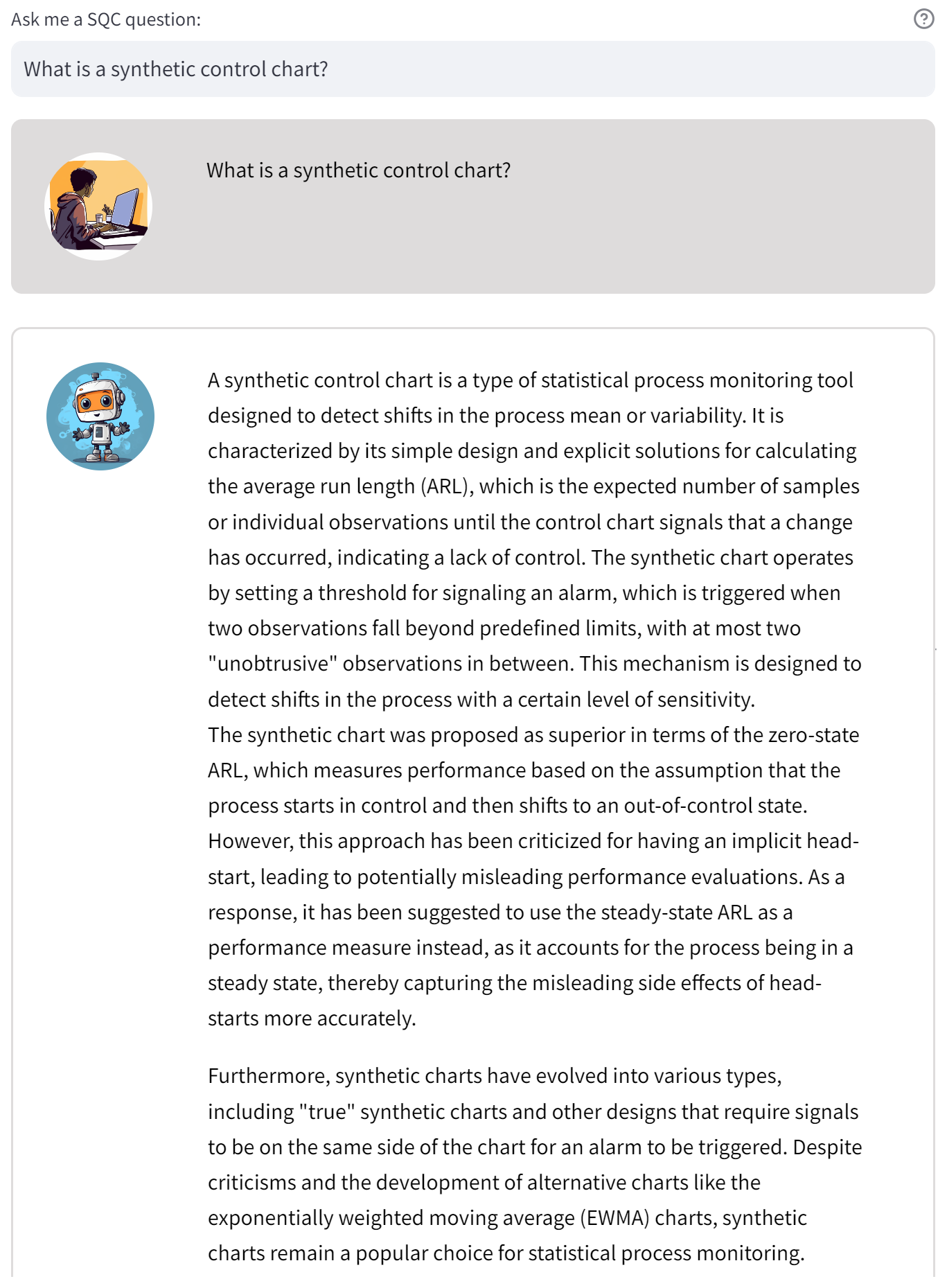}
         \subcaption[b]{\texttt{ChatSQC-Research}}
     \end{subfigure}
    \caption{Response of \texttt{ChatSQC} to Prompt 9.}
    \label{fig:prompt9}
\end{figure}

\begin{figure}[htb!]
    \centering
     \begin{subfigure}[T]{0.495\textwidth}
         \centering
         \includegraphics[width=\textwidth, trim={0 0 0 0.8in}, clip]{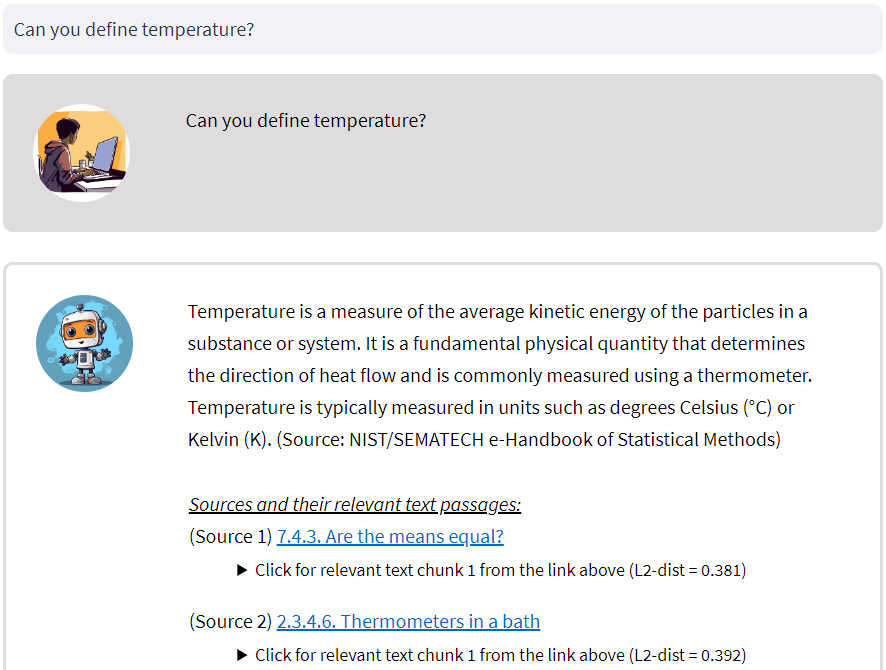}

        \vspace{\baselineskip}
         
         \subcaption[a]{\texttt{ChatSQC-Basic}}
     \end{subfigure}
     \begin{subfigure}[T]{0.495\textwidth}
         \centering
         \includegraphics[width=\textwidth, trim={0 0 0 1.4in}, clip]{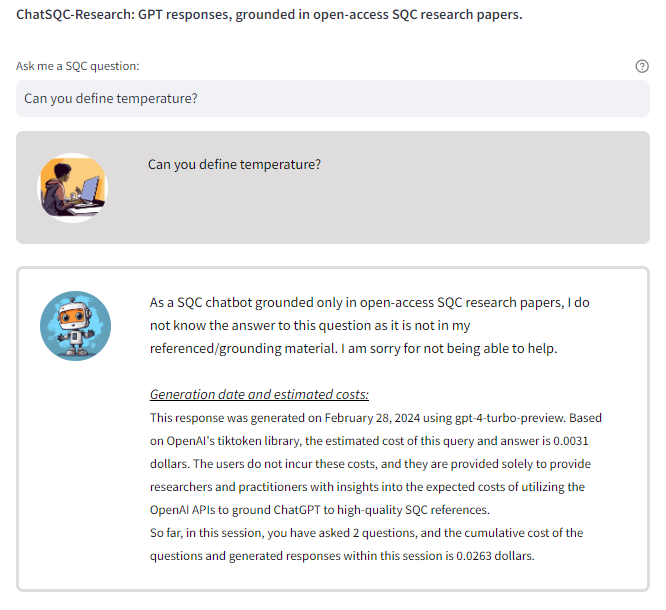}
         \subcaption[b]{\texttt{ChatSQC-Research}}
     \end{subfigure}
    \caption{Response of \texttt{ChatSQC} to Prompt 10.}
    \label{fig:prompt10}
\end{figure}

Here we examine how \texttt{ChatSQC-Basic} and \texttt{ChatSQC-Research} responded to prompts 9 and 10 (see Figures \ref{fig:prompt9} and \ref{fig:prompt10}). Prompt 9 asks the chatbots ``What is a synthetic control chart?"  As instructed when the topic of the question is not within its grounding material, \texttt{ChatSQC-Basic} correctly responds that this material is not within its grounding material (see Figure \ref{fig:prompt9}a).  \texttt{ChatSQC-Research}, on the other hand, provides a brief summary of the synthetic control chart. Synthetic control charts are discussed in \cite{knoth2022expanded} which is included in the grounding material for \texttt{ChatSQC-Research}.  

Prompt 10 asks the chatbots "Can you define temperature?"  This is a purposefully misleading question since "temperature" is a measure of kinetic energy, but also a parameter of LLMs that controls the variability in the responses.  Strictly speaking, the answer to this prompt is not in the NIST handbook upon which \texttt{ChatSQC-Basic} is grounded since there is no definition of temperature in the handbook. However, temperature is often used in the handbook as a factor in an experimental design. An early (August 2023) version of \texttt{ChatSQC-Basic} responded with a definition of temperature as a ``measure of kinetic energy of the particles in a substance or system." This answer, while correct, is outside of the grounding materials for \texttt{ChatSQC-Basic}. However, after updates to the LLM and our app, when entered now, the response from \texttt{ChatSQC-Basic} is that it does not know the answer because it is not in the reference material. \texttt{ChatSQC-Research}, which was developed later (January 2024), correctly responded that ``temperature" is not within its grounding materials. These results illustrate how the app safeguards against hallucinations when prompted outside of the scope of its grounding.

\subsection{\texttt{ChatSQC-Research} Responses}

In this subsection, we discuss the ratings of the responses to prompt 11 generated by \texttt{ChatSQC-Research} and the benchmark apps. We also include a twelfth ``paper'' in this discussion; this paper is a fictitious paper to study hallucinations. 

Table \ref{tab:descriptives11} presents the accuracy ratings and word counts for each of the three apps, averaged across the seven received ratings $(n=7)$. The table shows that the responses are similarly accurate, on average: \texttt{ChatSQC-Research} and \texttt{GPT-3.5} have very similar average accuracy values: 3.43 and 3.29, respectively. As expected, \texttt{GPT-4's} accuracy is slightly higher: 3.75. It should be noted that \texttt{GPT-4's} average accuracy is based on 4 out of the 7 responses. 

For three of the papers, \texttt{GPT-4} answered that it was unable to find the paper online and did not know what it was about. Interestingly, \texttt{GPT-3.5} did have an answer for those three papers. We have also included the average word count and see that our \texttt{ChatSQC-Research} provides shorter answers, on average, than \texttt{GPT-3.5} and \texttt{GPT-4 .}

The corresponding authors have also provided some comments giving insight into the responses provided by the apps. For our \texttt{ChatSQC-Research} some comments read:
\begin{itemize}[nosep]
    \item \emph{``the text is very strongly linked to our abstract and keywords. Of course, in general the abstract is also short summary according to us (the authors), so it makes sense that the algorithm picks up on that. "}
    \item ``\emph{it is accurate in the summary but very vague."}
    \item `` \emph{It lacks details about the methodology. The statement is too general. However, the description is not wrong."}
\end{itemize}
We see that in general \texttt{ChatSQC-Research} provides good, though slightly short, summaries. 

For \texttt{GPT-3.5} some of the comments read:
\begin{itemize}[nosep]
    \item \emph{``entirely accurate, a lot of details and all correct}"
    \item \emph{``More details are included. However, it is too verbose.}"
\end{itemize}
We see that a general sentiment is that \texttt{GPT-3.5} provides good, however slightly verbose summaries. Which is inline with the average word count which is highest for \texttt{GPT-3.5} (see Table \ref{tab:descriptives11}.) 

For \texttt{GPT-4} some of the comments  read
\begin{itemize}[nosep]
    \item \emph{``short but accurate in the description"}
    \item \emph{`` It is able to capture in few words the main intuition of the method."}
\end{itemize}

These are related to 4 out of the 7 papers for which \texttt{GPT-4} could find the paper online. However, for the papers that it could not find (3 out of 7) some comments read: 
\begin{itemize}[nosep]
    \item \emph{``It provides a general reference to the bootstrap method out of the scope of the paper."}
    \item \emph{`` Initially, it states that it cannot access/read the paper... but it does provide an educated guess of the contents based on the title. "}
\end{itemize}
This indicates that \texttt{GPT-4} ``guess" at the content of the paper based on the reference provided, as it is giving us an answer while explicitly stating that it does not have access to the paper it is answering about.

 \begin{table}[htb]
    \centering
    \caption{Descriptive statistics of the ratings for prompt 11.}
    \label{tab:descriptives11}
    \begin{tabular}{c ccc } 
    \toprule
     & \textbf{\textbf{ChatSQC-Research}} & \textbf{GPT-3.5}& \textbf{GPT-4}\\ 
    \midrule
    Average accuracy (std dev) & 3.43 (0.79)& 3.29 (1.11) & 3.75 (0.96)\\
    %Std Dev Accuracy & 0.81 & 1.05& 1\\
    Average word count (std dev)& 119.9 (60.1) & 192.0 (55.6)& 138.8 (48.6)\\
    %Std Dev word count& 11.7 & 57.3 & 18.5\\
    $n$       & 7& 7  &4 \\
    number of NAs & 0&   0&3 \\
    \bottomrule      
    \end{tabular}
\end{table}

\begin{figure}
    \centering
     \begin{subfigure}[t]{0.495\textwidth}
         \centering
         \includegraphics[width=.98\textwidth]{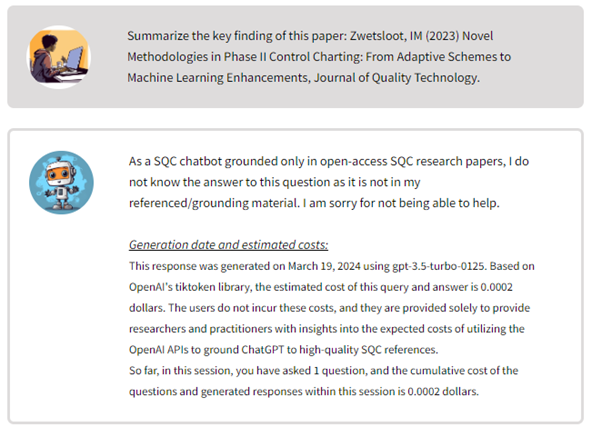}
         \subcaption[a]{\texttt{ChatSQC-Research}}
     \end{subfigure}
     \begin{subfigure}[t]{0.495\textwidth}
         \centering
         \includegraphics[width=.98\textwidth]{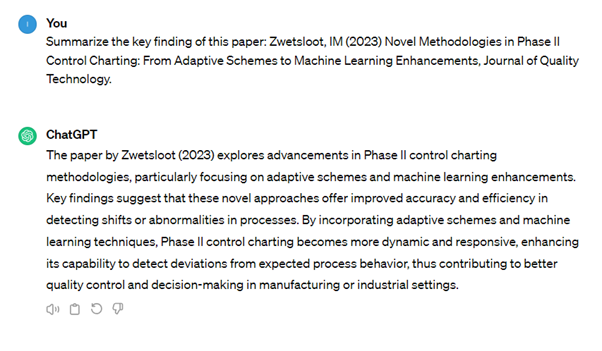}
         \subcaption[b]{\texttt{ChatGPT-3.5}}
     \end{subfigure}
     \begin{subfigure}[t]{0.495\textwidth}
         \centering
         \includegraphics[width=.98\textwidth]{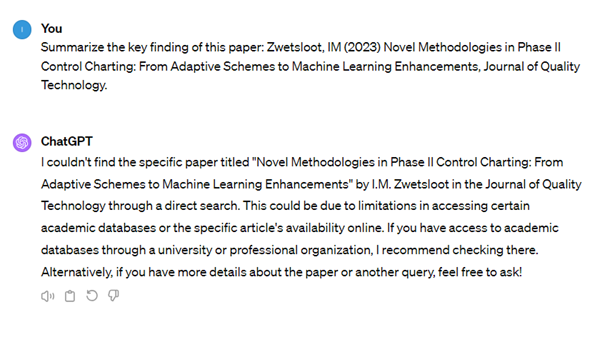}
         \subcaption[c]{\texttt{ChatGPT-4.0}}
     \end{subfigure}
    \caption{Response of apps to prompt 11 for a phantom paper}
    \label{fig:prompt11phantom}
\end{figure}

We have also asked each chatbot to summarize key findings of the fictitious paper ``Zwetsloot, IM (2023) Novel Methodologies in Phase II Control Charting: From Adaptive Schemes to Machine Learning Enhancements, Journal of Quality Technology.'' This paper does not exist, and the question was asked to test for hallucinations. A hallucination in LLMs occurs when an LLM generates false or partially true information based on extrapolations from its training material, logical errors, or both. Responses, shown in Figure \ref{fig:prompt11phantom}, correctly noted that this paper is not part of the grounding material of \texttt{ChatSQC-Research}.  Similarly, \texttt{GPT-4} states that it cannot find the paper. However, \texttt{GPT-3.5} summarizes this non-existent paper. This intentionally misleading question can be described as an adversarial attack on the model and gives insight into the limitations of the current \texttt{GPT-3.5} model.

Overall, we see that \texttt{ChatSQC-Research} provides the most conscise answers to prompts about research papers upon which it is grounded.  The answers are generally shorter than the ungrounded \texttt{ChatGPTs}, and the accuracy varies for all apps. We also note that the accuracy of \texttt{GPT-4} is the highest \emph{if it can find the paper (4 out of 7 in our experiment).} We note that the quality of the LLMs response is multi-dimensional and difficult to measure.  We are currently assessing this with expert accuracy ratings,  note that developing comprehensive measures to evaluate the quality of LLM responses is an important topic for future research. %Finally we have used \texttt{GPT-3.5} for the responses of \texttt{ChatSQC-Research}, using gpt 4 (which is a user choice) would also influence the results. It would be expected that both accuracy and wordcount would increase. }

\subsection{Overall Evaluation}

Overall, the accuracy of our \texttt{ChatSQC} (Basic and Research) is comparable to the benchmark methods, though the restricted content of the included referenced sources sometimes limits the accuracy. A large benefit of our app is that it can respond that it does not know the answer to a prompt when there is no context for it in the training material. This was not the case with the ungrounded ChatGPT-3.5. Furthermore, our ChatSQC app provides the relevant references used in its retrieval augmented generation, along with their most pertinent text chunks, providing users with the ability to assess whether the answer is truly from its knowledge base or that it may be the result of hallucinations or unverified external information that is available to the LLM. Note that we have only studied accuracy; in the future, it would be interesting to look at other aspects of the responses, such as conciseness, depth, and usefulness.

%% file: 05_discussion.tex
\instructions{
\begin{enumerate}[label=(\arabic*)]
    \item This subsection should discuss that it is nice that we can augment LLMs with seminal work or more recent references. It should be noted that while it is possible to include documents in ChatGPT (with a code interpreter, for example), this is different. It provides three advantages: (a) learning by doing (using better embeddings, comparing and contrasting different LLMs, a playground for DoE and quality technology research); (b) data privacy, as ``OpenAI will not use data submitted by customers via our API to train or improve our models unless you explicitly decide to share your data with us for this purpose." \href{https://openai.com/policies/api-data-usage-policies}{API data usage policies}; and (c) can also be hosted locally, which can provide greater reliability and removes current throttling limits set by OpenAI through their ChatGPT app.

    \item A subsection about the research opportunities that this offers to our community (but some of them are about learning by doing).

    \item A Guide for Practitioners 

    \item Limitations of our app, emphasizing that it is in its Alpha stage of software development. 
\end{enumerate}
}

\subsection{The Use of APIs and Grounding to Augment Existing LLMs}

This paper provides a template for how specific SQC references can ground existing LLMs. We note that AI-based generative chatbots, for example, ChatGPT (GPT-4 with code interpreter that is currently available to ChatGPT Plus subscribers) and \href{https://claude.ai/}{claude.ai} (by Anthropic), can be used to analyze documents such as PDF, HTML, and CSV files. However, our approach is different.  We limit the chatbot's answers to only information in our referenced material; thus, we obtained more precise responses for SQC-explanatory prompts than \citet{megahed2023generative}. In addition, we reduce the risk of hallucinations and provide explicit source references. The use of APIs through the \texttt{langchain} Python library presents the following advantages over the ChatGPT app/web interface:
\begin{enumerate}[nosep, label = (\arabic*)]
    \item \textbf{Enhanced Data Privacy and Performance:} Focusing on the current alpha version using Open AI's APIs, our app provides better (defaults) for data privacy when compared to the web interface for ChatGPT since ``OpenAI will not use data submitted by customers via our API to train or improve our models unless you explicitly decide to share your data with us for this purpose." \citep{openai2023data} Furthermore, the rate limits are much higher for the API than for the web interface, so SQC users are less likely to be throttled / rate limited. 
    
    \item \textbf{Flexibility and Cost Efficiency:} Minimal changes are needed to use a different LLM in the code base. In our alpha version, we have used the \texttt{GPT-3.5 Turbo} and \texttt{GPT-4 Turbo} models. However, in future versions, we may use (a) a locally-hosted and free open-source model such as \texttt{Llama-2-7B-chat} \citep{touvron2023llama2}, or (b) a model with a much larger context window, e.g., \texttt{Gemini Pro 1.5} \citep{pichai2024gemini}, where we will explore the utility of putting the entire referenced materials in the context window.
    % \begin{enumerate}[nosep, label = (\alph*), leftmargin=*]
    %     \item is not a black box since the model's code, weights, and parameters are publicly available;
    %     \item can be hosted locally, without transferring embeddings/prompts to third parties; and
    %     \item cost efficient as we will not be paying API charges for the embedding and GPT APIs. 
    % \end{enumerate}
    \item \textbf{Understanding the Intricacies of LLMs:} By grounding ChatGPT with SQC knowledge, we gained a much better understanding of the workings of LLMs. Sharing our experience/knowledge is valuable to both researchers and practitioners as they embark on their own journey to use LLM as virtual assistants.
\end{enumerate}

\subsection{ChatSQC's Implications to SQC Researchers}

With our provided source code, SQC researchers can implement their own playground/test-bed to further advance the use of generative AI in SQC research. In the following paragraphs, we highlight five potential SQC-related research opportunities.

Firstly, creating \texttt{ChatSQC} offers a robust testing ground for SQC researchers. With our provided source code, SQC researchers can perform numerical experiments to test the impact of input parameter choice on the generated output quality. The input parameters include but are not limited to, the following: (a) the choice of document readers can differ based on the journal's formatting and manuscript contents (e.g., we used the `PyMuPDFLoader()' method from the \texttt{langchain} library to read the 52 research papers as it presented the best results in our initial experiments; however, quantifying whether our initial observations are correct and whether it holds for a larger pool of journals remain open questions); (b) how the text is chunked (i.e., text chunk size and the amount of overlap between successive text chunks) since this affects the app's ability to understand and process information; (c) the choice of embedding models, places semantically similar words/tokens close together in the embedding space, enhancing the relevance of search and retrieval operations; (d) vector stores, where the embedding vectors are stored (functionalities and how text similarities are captured can be different across the different vector store databases, which can impact the efficiency and accuracy of retrieving similar texts); and (e) the choice of the LLM model, determines the overall capability of the app to generate high-quality, contextually relevant responses. A structured experimental design can be used to examine how these choices contribute to the quality of the generated text and whether the contribution is consistent for different LLM tasks (e.g., information retrieval, coding [i.e., translation from natural language to machine code], and summarization). In the context of machine learning, the work of \citet{ahady2023explaining} can serve as a starting point for this stream of research. In addition, from a measurement theory viewpoint, researchers can delve into various methods to define and measure the quality and variability of the generated outputs. This dual exploration of experimental design and measurement theory can enrich the generative AI and SQC/quality control literature.

Secondly, there are promising implications if our approach to grounding LLMs is extended to a larger collection of SQC research. For example, consider the scenario where copyright permissions are obtained for all articles in ASQ research journals. If this is possible, a more advanced version of \texttt{ChatSQC-Research} can be developed. It would be interesting to examine whether a better-grounded version of \texttt{ChatSQC-Research} can revolutionize the way we conduct literature reviews, identify gaps in existing research, and facilitate replication of simulation and numeric methods, even when published without the original code. The current version of the \texttt{ChatSQC-Research} seems to be promising in performing Q\&A on the referenced papers. However, RAG-based approaches may not be needed in the future if existing/future LLMs continue with the trends of more oversized context windows and complementary inexpensive API usage costs. For example, the Gemini 1.5 Pro (via AI Studio) can accept up to 1,000,000 tokens \citep{pichai2024gemini}, i.e., approximately 1,000--1,250 pages of text. This means that we can conceivably put the entire collection of our reference materials as part of the prompt for a future iteration of \texttt{ChatSQC-Research} and directly ask questions about these materials. This will likely improve the LLM's ability to synthesize the material and potentially identify gaps/opportunities in the literature.  %The recent publication of \texttt{LongNet} \citep{ding2023longnet}, which scales transformers to model long sequences of up to a billion tokens, \rev{would even make this task computationally possible for the entire collection of leading SQC journals. A potential open question would be whether the improved context and higher expected quality of the generated text would outweigh the latency in response and increased cost (we are currently tokenizing approximately 1,500 tokens for both the prompt and response).} 

A third avenue of research involves the examination of the impact of a mix of high-quality and lower-quality references on the performance of a tool such as in our proposed \texttt{ChatSQC-Research} bot. In the context of general LLMs, they are trained to predict the next word, and then they are typically fine-tuned using methods such as Reinforcement Learning from Human Feedback (RHLF). In the context of grounding LLMs with additional materials/references, we can examine the impact of reference quality on the performance of the LLM. For instance, what happens when substandard methodologies, such as the refuted triple exponential smoothing charts \citep{knoth2023critique}, are incorporated into the training material? Do these lower-quality methodologies influence the bot's performance, and if so, to what extent? Determining the threshold for such influences is an intriguing research question, which can provide information on how LLMs ``behave'' and potentially be useful when developing other domain-specific bots. %For example, GPT-4 was shown to have unwanted skills, such as teaching potential threat actors how to prepare biological weapons, which OpenAI had to spend a significant effort trying to prevent \citep{openai2023gpt4}. Concurrently, exploring a crowd-sourcing mechanism to infuse the bot with diverse, yet scientifically valid, research ideas might prove beneficial.

Fourthly, there is a need to continually monitor LLM quality over time. Black-box models, e.g., GPT 3.5 and GPT 4.0, require attention to how companies' (such as OpenAI) tweaks to a given base model can impact its performance over time. For example, in a preprint, \citet[p. 1]{chen2023chatgpts} noted that ``GPT-4 (March 2023) was very good at identifying prime numbers (accuracy 97.6\%) but GPT-4 (June 2023) was very poor on these same questions (accuracy 2.4\%). Interestingly GPT-3.5 (June 2023) was much better than GPT-3.5 (March 2023) in this task. GPT-4 was less willing to answer sensitive questions in June than in March, and both GPT-4 and GPT-3.5 had more formatting mistakes in code generation in June than in March. Overall, our findings show that the behavior of the same LLM service can change substantially in a relatively short amount of time, highlighting the need for continuous monitoring of LLM quality.'' %Without going into details on how OpenAI responded to these accusations/findings or their validity, we agree that there is a need for continuous monitoring of LLM quality over time. 
Developing methods to monitor LLMs is a research opportunity that seems natural to our SQC community.  

Finally, studying how to measure LLM output quality seems a natural fit for our community.  \citet[p.7]{bowman2023things} noted that ``brief interactions with LLMs are often misleading... This instruction-following behavior is not inherent to the model but rather is grafted onto it using highly imperfect tools... In part because of this, models can be sensitive to the contents of their instructions in idiosyncratic ways. Often, a model will fail to complete a task when asked, but will then perform the task correctly once the request is reworded or reframed slightly, leading to the emerging craft of \textit{prompt engineering}.'' %From a quality measurement and experimental design perspective, examining how this phenomenon can be accounted for would be interesting. Although the concept of \textit{replications} will likely play a role, 
We hope that the SQC and experimental design community can collaborate with measurement theory experts to incorporate output metrics that are more informative than the existing \texttt{pass@k} metric \citep{kulal2019spoc,chen2021evaluating}, where $k$ samples are generated for a given prompt/problem and the solution is considered correct if any sample passes the unit tests. 
%\info{The website \url{https://midas.umich.edu/generative-ai-resources/} curates example Research Uses of Generative AI, we can possibly provide highlight its existence to our community.}

\subsection{ChatSQC's Potential Impact on Teaching and Learning}
In general, LLMs (with plugins such as Code Interpreter) will contribute to a new paradigm of teaching and learning. Students can now use LLMs to quickly generate/explain code and analyze datasets. In our opinion, these LLM abilities should transform the educational landscape, where universities and employers need to forecast the skills needed for future knowledge workers. For example, many statistics and data science courses emphasized how to code/implement taught methods as part of the expected learning outcomes. With the ability to quickly generate (often working) code, it is unclear how existing courses will be transformed. Coding logic will likely remain an important consideration, at least for the near future, so that generated code can be evaluated, debugged, and optimized. However, we estimate that students will have to now learn about simulating datasets and writing minimal, privacy-preserved, examples to protect their future employers' intellectual property while capitalizing on LLMs to boost their productivity/ performance.     

With \texttt{ChatSQC} (and similar grounded/augmented LLMs), teachers can provide student learners with a dedicated virtual assistant whose answers can be solely based on selected textbooks and reference materials. %\citet{mollick2023assigning} identified several use cases for LLM-based virtual assistants, highlighting their potential pedagogical benefits and risks. In our estimation, the grounding of \texttt{ChatSQC} allows it to serve as a: (a) mentor, providing frequent feedback to improve learning outcomes, even when the student does not take all its advice; (b) tutor, providing personalized instruction to students; (c) teammate, providing alternate definitions/viewpoints based on its grounding material, and (d) tool, assisting students with accomplishing more within the same time frame. The interested reader is referred to \citet{mollick2023assigning} for more details. 
%It is important to note that 
The \texttt{ChatSQC} app can be grounded with additional material by teachers. This can include other textbooks, or select reading materials. The ethical use of potentially copyrighted materials must be considered in such a case. We used a government-backed, public-domain textbook in our current version of \texttt{ChatSQC}. However, in the case of the adaption of \texttt{ChatSQC} to include copyrighted materials, we envision two possible scenarios for a tailored/customized implementation of our \texttt{ChatSQC}-like tool. First, we suspect that students will have \texttt{ChatSQC}-like chatbots running locally on their machines. This is possible with lightweight and free embeddings and LLMs such as \texttt{GPT4All} \citep{gpt4all}, which do not require an Internet connection or graphical processing units (GPU). %This case will likely involve students using their purchased electronic copies of their textbook to ground the chatbot, which: (a) can be easily implemented with minor modifications to our source code, and (b) will likely meet the U.S.' \texttt{Fair Use} doctrine of copyrighted materials since the students will not be commercializing/publishing their tool, and they would have already purchased the copyrighted material. We note that the fair use of copyrighted materials can differ by country. 
Second, publishers may provide \texttt{ChatSQC}-like chatbots as complementary materials to their copyright-protected textbooks. %We are currently in talks with one of the textbook publishers to construct a \texttt{ChatSQC} bot, grounded in a popular quality control textbook. 
In either case, the students will have access to a chatbot that can quickly answer (some of) their questions, and they will have to learn to verify these answers. Furthermore, these bots can provide students with machine translations of their reference text to their native language (e.g., we have tested \texttt{ChatSQC}'s ability to produce responses in Arabic, Chinese, Dutch, French, German, and Spanish). %In coding examples, bots can potentially show students how code can be translated into other programming languages. 

\subsection{A Guide to Practitioners}

\texttt{ChatSQC}, in its current form, provides practitioners with a versatile tool that can be leveraged in \ul{three} ways, catering to diverse needs and use cases within industrial statistics, quality control, reliability, and experimental design. Firstly, practitioners can directly interact with our hosted \texttt{ChatSQC} bot. This mode of use provides quick access to definitions and explanations of concepts, helping practitioners navigate complex terminologies and understand key ideas. %This usage is particularly beneficial for those seeking a readily available and easy-to-use resource to enhance their knowledge and understanding of the field. 
Secondly, for organizations desiring more control over the reference material used, the option to host their version of \texttt{ChatSQC} exists. Our source code allows for including PDF or HTML documents as references, offering flexibility in the choice of augmenting material. Companies can use PDFs of different books and research papers internally, provided they have purchased the materials (and possibly their copyrights), tailoring the tool to their unique requirements and preferences. Lastly, practitioners can use \texttt{ChatSQC} as a foundation for more extensive applications within their organizations. The tool could be adapted to synthesize information from various unstructured text within the organization, such as maintenance manuals, warranty claims, quality, and/or Lean Six Sigma procedures. This advanced use of \texttt{ChatSQC} could provide organizations with a customized tool for internal knowledge management and information retrieval. %These diverse usage possibilities make \texttt{ChatSQC} a valuable tool for practitioners, offering both ready-to-use benefits and the potential for customization and expansion to meet specific organizational needs.

Looking forward, chatbots like \texttt{ChatSQC} have the potential to shape the future of work in industrial statistics, quality control, reliability, and experimental design. %As automation and digitization continue to transform workplaces, tools like \texttt{ChatSQC} could play a pivotal role in facilitating this transition. 
By automating access to complex terminologies and explanations, \texttt{ChatSQC} could reduce the time spent accessing relevant reference materials and allow practitioners to focus more on strategic decision-making and innovative problem-solving. The ability to customize and expand the tool and integrate it with other data sources within an organization opens up possibilities for developing comprehensive AI-driven solutions. \texttt{ChatSQC} provides an example of how LLMs can manage and retrieve information in our field.  %These could revolutionize how information is managed and utilized in our field, driving operational efficiencies and fostering a culture of continuous learning and adaptation (consistent with widely adopted six sigma, lean manufacturing, and/or statistical engineering principles). 
%\info{The table in \url{https://midas.umich.edu/generative-ai-resources/} might be worth highlighting to practitioners; not sure of the exact location we should cite it.}

\subsection{Limitations}

In addition to the specific challenges we have encountered in developing \texttt{ChatSQC}, broader limitations and obstacles should be acknowledged in the study of large language models (LLMs). Much of the research concerning LLMs, including many pioneering studies that introduced novel methods or theories, is not published in peer-reviewed venues \citep{bowman2023things}. Additionally, some LLM providers, such as OpenAI, have treated the details of their LLM design and training as propriety information, which is an obstacle to scientific progress. ``This means that surprising novel claims about LLMs are often the product of messy, fallible science that goes beyond established disciplinary practice.'' \citep[p. 9]{bowman2023things}

Turning to the specific limitations of our \texttt{ChatSQC} bot, it is essential to note that our current tool is at an alpha stage, serving as a proof-of-concept. As such, several limitations intrinsic to this stage of development must be considered.
\begin{enumerate}[label = (\arabic*), nosep]
    \item \textbf{Resource Dependence:} The performance of our grounded ChatGPT is dependent on external resources, specifically the APIs used in its development and operation. Each interaction with the ChatGPT involves API calls, which incur costs. Consequently, the scale of deployment and usage of our ChatGPT can be constrained by the available financial resources, potentially limiting its accessibility and performance. In Appendix \ref{cost:emb}, we have provided a detailed and conservative estimate of the cost of the \texttt{ChatSQC} app.
    
    \item \textbf{Narrow Focus and Depth Limitations:} The bot operates proficiently within the SQC domain, but its depth of understanding and ability to handle topics beyond this scope is limited. Furthermore, the bot's understanding and responses are strictly bound by the contents of the referenced material. Suppose certain concepts related to SQC are absent or only briefly introduced in the reference book. In that case, the bot may not fully capture the nuances that require a more thorough examination or understanding (for example, prompt 6, as discussed in Section 3). This limitation could affect the responses' depth, potentially hindering the bot's usefulness for tackling complex or nonstandard SQC problems.

    \item \textbf{User Interaction:} The users' interaction with \texttt{ChatSQC} might not always yield the desired results. The intricacies of \textit{prompt engineering}, the precise wording and framing of queries, significantly influence the effectiveness of user interactions with the bot. This might necessitate users to acquire certain skills in framing their queries.

    \item \textbf{Maintaining Relevance:} Keeping our \texttt{ChatSQC} current in the face of rapid advancements in SQC methodologies can be challenging. Incorporating new methodologies and approaches would require regular updates and retraining, a process that might not always be feasible due to resource constraints.
\end{enumerate}
Despite these limitations, we believe that by releasing this alpha version and encouraging user interactions and feedback through our GitHub repository, we can harness valuable insights to address the \texttt{ChatSQC}-specific issues. The input from users will provide us with a real-world understanding of how the bot is being used, its current limitations, and areas of potential improvement, enabling us to refine and enhance future versions.

%% file: 06_conclusions.tex
%To say there is a current gold rush to understand the nature and use of LLMs, AI, and AI-augmented work is an understatement. In our opinion, 
AI-augmentation will continue to radically transform the nature of education and work.  In the context of developing the \texttt{ChatSQC} app, for example, we learned of several rapid AI integrations.  Stack Overflow has introduced generative AI into their public platform, Stack OverflowAI \citep{chandrasekar2023overflowai}, which includes enhanced search, and integration with Salesforce's Slack platform and Microsoft's Visual Studio Integrated Development Environment (IDE). We used the LangChain framework to develop our \texttt{ChatSQC} app.  During the development of the \texttt{ChatSQC} app, the LangChain documentation added enhanced search capability through the commercially available Mendable product (\url{https://www.mendable.ai/}).  Mendable employs LLM search capability specifically trained on websites, customer support documentation, sales information, and many other sources.  

These are just a few examples of how generative AI technology is rapidly changing the management and procurement of knowledge.  As with any new technology, there are many open challenges, risks, and potentials for exploitation.  Currently many are simply exploring the utility and ethical use of such systems.  For example, \citet{bender2021stochastic} discussed the benefits, and risks, and questioned the necessity of such models.   \citet{mitchell2023how} in their \textit{Science} expert voices article states ``The assumptions that we make for humans—that they cannot memorize vast collections of text related to test questions, and when they answer questions correctly, they will be able to generalize that understanding to new situations—are not yet appropriate for AI systems.''  Others believe that LLMs have the capability to not only parrot responses but to actually understand the world.  For example, with the development of \texttt{Othello-GPT} \citep{li2023emergent}, the authors showed that LLMs are not just ``stochastic parrots'' \citep{bender2021stochastic}; instead, LLMs can create a correct model related to the Othello game by being trained on the moves and not the rules of the game.  Some experts, including Andrew Ng and Geoff Hinton, agree that LLMs have the capability of some level of understanding \citep{ng2023conversation}.    

Our work on this paper and app arose from a conversation among colleagues on the potential power of generative AI in the field of SQC \textbf{if the model is grounded in accurate and reliable content}.  The Alpha version of \texttt{ChatSQC} is our first attempt at introducing a specifically grounded generative AI chatbot that can broaden the exposure of sound SQC resources to a larger community. By providing a dedicated web platform along with the code used to generate the app, we aim to encourage others in the SQC community to contribute to the training content of \texttt{ChatSQC}. Our \texttt{ChatSQC} app is committed to continuous improvement, ensuring it remains updated with the latest state-of-the-art embedding and GPT models, a strategy that guarantees the app maintains optimal performance. We believe that \texttt{ChatSQC} is a clear testament to the transformative potential of generative AI in our community, and we hope to spur advancement in the integration of SQC knowledge and AI technology.   

%% file: 07a_code.tex
To encourage extensions to our app, we made the code used for the app publicly available on GitHub with an MIT license. Our GitHub repository can be accessed at 
\url{https://github.com/fmegahed/chatsqc}; it contains the code used to scrape and preprocess our reference material and the Python code for the chatbot's graphical user interface along with the CSS styling files. It is important to note that we do not provide our OpenAI API key on the GitHub repo and that people interested in running our app (locally) or extending it should obtain an API key from \url{https://platform.openai.com/account/api-keys}. Our \texttt{ChatSQC.py} file assumes that there is an ``.env'' file, which contains a variable called \texttt{OPENAI\_API\_KEY}, where you can assign your actual API Key as: \texttt{OPENAI\_API\_KEY}$=$\texttt{sk-mo9KXYZfk7pvRnIcdZzPFU8WlzuJB1EFLmihGYop4YZnTjk}. Note that this is a dummy API key, where we have maintained the format of the real key, but it should not have any real functionality or access permissions. Please refer to our Repo's \url{https://github.com/fmegahed/chatsqc/#readme} for more details.

%% file: 07b_verifiability.tex
In the spirit of scientific rigor and transparency, we present in this section an approach tailored to bridge the inherent limitations of repeatability associated with Large Language Model (LLM) experimentation. In an effort to navigate the unique challenges posed by these models, we have compiled a suite of comprehensive visual materials that serve to validate our research findings and provide a tangible record of our methodology. We have designated this subsection as ``Interactive Verification", which we encourage other researchers to use as a guide for their future submissions of LLM-related research papers. It contains videos and images documenting every interaction we have conducted with the LLM. In doing so, our objective is to provide a visually verifiable trail, to encourage scrutiny, promote transparency, and stimulate further discourse in this dynamic field of research. In this light, we invite our peers to review this evidence, observe the LLM in operation, and draw their own conclusions based on the empirical data presented below.

\begin{itemize}[nosep]
\item \textbf{CSV file of accuracy ratings for prompts 1-8}: \url{https://github.com/fmegahed/chatsqc/blob/main/accuracy%20experiments/rating_accuracy.csv} 
\item \textbf{CSV file of accuracy ratings for prompt 11:} \url{https://github.com/fmegahed/chatsqc/blob/main/accuracy%20experiments/ratings_experiment2_blinded.csv} 
\item \textbf{\texttt{ChatSQC-Basic} screenshots for prompts 1-10:} We provide screenshots of the prompts and their responses at: \url{https://github.com/fmegahed/chatsqc/tree/main/accuracy%20experiments/ChatSQC_screenshot}
\item \textbf{GPT-3.5 screenshots:} We provide screenshots of prompts 1-8 and their responses at: \url{https://github.com/fmegahed/chatsqc/tree/main/accuracy%20experiments/GPT-3.5%20screenshots}
\item \textbf{GPT-4 screenshots:} We provide screenshots of prompts 1-8 and their responses at:  \newline  \url{https://github.com/fmegahed/chatsqc/tree/main/accuracy%20experiments/GPT-4%20screenshots}
\item \textbf{Prompt 11:} We provide screenshots of prompt 11 for 11 articles and the responses at:  \newline  \url{https://github.com/fmegahed/chatsqc/tree/main/accuracy%20experiments/prompt11_LLMresponses}
\end{itemize}

%% file: 07c_app_articles.tex
% \vspace{-10pt} 

\begin{table}[htb!]
\caption{Journal Articles used to Ground \texttt{ChatSQC-Research}.}
\centering
\begin{tblr}{
  width = \linewidth,
  colspec = {Q[100]Q[45]Q[65]Q[200]},
  row{1} = {font=\bfseries}, 
  cell{2}{1} = {r=7}{},
  cell{2}{2} = {r=7}{},
  cell{9}{1} = {r=4}{},
  cell{9}{2} = {r=3}{},
  cell{13}{1} = {r=13}{},
  cell{13}{2} = {r=8}{},
  cell{21}{2} = {r=5}{},
  vline{2-4} = {1-26}{},
  vline{4} = {3-8,10-11,14-20,22-25}{},
  vline{3-4} = {12,21}{},
  hline{1-2,9,13,26} = {-}{},
  hline{3-8,10-11,14-20,22-25} = {3-4}{},
  hline{12,21} = {2-4}{},
}
Journal & License & Field & Papers\\
Technometrics  & BY & Bayes & \citeappx{yang2023decision, rougier2023bayesian, yannotty2023model} \\
    &   & change point & \citeappx{bardwell2019most} \\
    &   & classification    & \citeappx{raymaekers2022class} \\
    &   & computer experiments & \citeappx{oakley2017calibration} \\
    &   & exp design           & \citeappx{overstall2017bayesian} \\
    &   & fda                  & \citeappx{hullait2021robust}* \\
    &   & spc                  & \citeappx{ryan2023detecting}* \\
Quality Engineering   & BY    & reliability          & \citeappx{zhou2023data}* \\
    &         & several              & \citeappx{capaci2019revised}*\\
    &         & spc                  & \citeappx{diko2017phase}, \citeappx{kuiper2023optimized}*, \citeappx{zwetsloot2017head}\\
    & BY-NC   & sampling             & \citeappx{luca2020web} \\
Quality and Reliability Engineering International & BY      & AI  & \citeappx{giudici2022explainable} \\
    &         & ML                   & \citeappx{cacciarelli2023robust, maculotti2024optimisation, wang2023three} \\
    &         & change point         & \citeappx{chapman2020assessment} \\
    &         & exp design           & \citeappx{otava2020communicating, alomair2021three}\\
    &         & image processing     & \citeappx{maculotti2022gaussian}\\
    &         & msa                  & \citeappx{celano2022bootstrap}* \\
    &         & reliability          & \citeappx{rabhi2023discrimination, bibartiu2024availability, amran2023critical, hijazy2020optimal, balakrishnan2023robust, leckey2020prediction, soleimani2021diagnostics} \\
    &         & spc                  & \citeappx{,abbasi2021efficient,capaci2020monitoring}, \citeappx{capezza2022functional}*,\citeappx{does2020design}*, \citeappx{knoth2023another, kostyszyn2021statistical, jones2023novel}, \citeappx{mahmood2021efficient}*, \citeappx{martinez2020one, nassar2021semiparametric,ottenstreuer2021control} \\
    & BY-NC   & ML                   & \citeappx{rotari2023variable}* \\
    &         & computer experiments & \citeappx{meynaoui2023second} \\
    &         & image processing     & \citeappx{kirchhoff2020detection}\\
    &         & reliability          & \citeappx{li2021remaining, zhao2023reliability} \\
    &         & spc                  & \citeappx{huberts2022predictive,  knoth2022expanded1}, \citeappx{ottenstreuer2022shiryaev}*, \citeappx{von2024comparison, weiss2022monitoring}  
\end{tblr}
\footnotesize{*Paper included in evaluation study of \texttt{ChatSQC-Research}}
\end{table}

%% file: 07d_cost.tex
We estimate the cost of embeddings and for processing queries individually below. 

\label{cost:emb}
\noindent \textbf{Assumptions for Computing the Cost of Embeddings:}
\begin{itemize}[nosep]
    \item OpenAI's explanation of a token: \href{https://openai.com/pricing}{``You can think of tokens as pieces of words, where 1,000 tokens are about 750 words.''}
    \item Cost per 1,000 tokens $=$ \$0.00010 (\textbf{PS:} cost for ADA v2 per the latest OpenAI pricing)
    \item Average length of a journal paper $=$ 25 pages
    \item Number of journal papers $=$ 52
    \item Average tokens per page $=$ 1000 (\textbf{PS:} this is a conservative estimate (likely around 700))
\end{itemize}

\vspace{.5\baselineskip}

\noindent \textbf{The total cost can be calculated as follows:}

\vspace{-\baselineskip}

\begin{align*}
\text{Total Tokens} &= \text{Avg Length of Paper} \times \text{\# of Journal Papers} \times \text{Avg Tokens per Page} \\
                    &= 25 \times 52 \times 1000 \\
                    &= 1,300,000 \text{ tokens}
\end{align*}

\vspace{-\baselineskip}

\begin{align*}
\text{Total Cost} &= \frac{\text{Total Tokens}}{1,000} \times \text{Cost per 1,000 Tokens} \\
                  &= \frac{1,300,000}{1,000} \times 0.00010 \\
                  &= \$0.13
\end{align*}

\vspace{-0.5\baselineskip}

\noindent Therefore, the total cost to embed 52 journal papers at an average length of 25 pages using the ADA v2 embedding model is \$0.13. We can easily rerun the embeddings each month (with additional papers) even if the number of open-access papers doubles over the next 3-5 years.

\vspace{\baselineskip}

\label{cost:queries}
\noindent \textbf{Given the following assumptions for processing queries:}
\begin{itemize}[nosep]
    \item OpenAI's explanation of a token: \href{https://openai.com/pricing}{``You can think of tokens as pieces of words, where 1,000 tokens are about 750 words.''}
    \item Average token size for the input query $=$ 250 tokens (\textbf{PS:} conservative assumption of an input query of $\approx$ 188 words)
    \item Average token size for the output = 1,000 tokens (\textbf{PS:} conservative avg output query length of $\approx$ 750 words)
    \item Number of input queries per month = 2,000
    \item Number of output queries per month = 2,000
    \item Input token cost = \$0.01 / 1,000 tokens (\textbf{PS:} OpenAI's latest pricing for GPT-4 Turbo)
    \item Output token cost = \$0.03 / 1,000 tokens (\textbf{PS:} OpenAI's latest pricing for GPT-4 Turbo)
\end{itemize}

\vspace{\baselineskip}

\noindent \textbf{The total cost for processing queries can be calculated as follows:}

\vspace{-\baselineskip}

\begin{align*}
\text{Total Input Tokens} &= \text{Number of Input Queries} \times \text{Average Input Query Length} \\
                          &= 2,000 \times 250 \\
                          &= 500,000 \text{ tokens}
\end{align*}

\vspace{-\baselineskip}

\begin{align*}
\text{Total Output Tokens} &= \text{Number of Output Queries} \times \text{Average Output Query Length} \\
                           &= 2,000 \times 1,000 \\
                           &= 2,000,000 \text{ tokens}
\end{align*}

\vspace{-\baselineskip}

\begin{align*}
\text{Total Input Cost} &= \frac{\text{Total Input Tokens}}{1,000} \times \text{Input Token Cost} \\
                        &= \frac{500,000}{1,000} \times 0.01 \\
                        &= \$5.00
\end{align*}

\vspace{-\baselineskip}

\begin{align*}
\text{Total Output Cost} &= \frac{\text{Total Output Tokens}}{1,000} \times \text{Output Token Cost} \\
                         &= \frac{2,000,000}{1,000} \times 0.03 \\
                         &= \$60.00
\end{align*}

\noindent Therefore, the total running cost for handling 2,000 input and 2,000 output queries, with the given assumptions, is \$60.00 per month. We can easily utilize our institutional support to pay for these costs. 

The issue becomes if our ChatSQC becomes very popular, with 8,000+ input and output queries per month, we will then attempt to democratize access to our app via (a) asking the community to host multiple mirrors or (b) asking each user to provide their OpenAI key. For a proof-of-concept, this would be a good problem to have if we reach this type of usage level.